\documentclass[10pt,sigconf,letterpaper,nonacm]{acmart}

\def\BibTeX{{\rm B\kern-.05em{\sc i\kern-.025em b}\kern-.08emT\kern-.1667em\lower.7ex\hbox{E}\kern-.125emX}}
    
\usepackage{hyperref}
\usepackage{algorithmic}
\usepackage{wasysym}
\usepackage{graphicx}
\usepackage{textcomp}
\usepackage{xcolor}
\usepackage{colortbl}
\usepackage{underscore}
\usepackage{multirow}
\usepackage{array}
\usepackage{balance}
\usepackage[normalem]{ulem}
\usepackage{comment}

\usepackage{subfigure}

\usepackage{fancyhdr}
\usepackage{extramarks}

\usepackage[autostyle]{csquotes}

\newcommand{\revPAM}[2]{#2}
\newcommand{\cameraPAM}[2]{#2}

\newcommand{\surveyPAM}[2]{#2}

\newcommand{\extraPAM}[2]{#2}

\begin{document}
\title{Revocation Statuses on the Internet}
%
%
%
%

\author{Nikita Korzhitskii}
\affiliation{%
  \institution{Link\"oping University, Sweden}
    \country{}
}

\author{Niklas Carlsson}
\affiliation{%
  \institution{Link\"oping University, Sweden}
  \country{}
}





%
        


\begin{abstract}

The modern Internet is highly dependent on the trust communicated via X.509 certificates.  However, in some cases certificates become untrusted and it is necessary to revoke them.  In practice, the problem of secure certificate revocation has not yet been solved, and \revPAM{there are today no}{today no} revocation procedure (similar to Certificate Transparency w.r.t. certificate issuance) \revPAM{}{has been adopted to} provide transparent and\revPAM{/or}{} immutable \revPAM{revocation history}{history} \revPAM{for}{of} all revocations\revPAM{e.g., like what}{}.  Instead, the status of most certificates can only be checked with Online Certificate Status Protocol (OCSP) \revPAM{or downloading a}{and/or} Certificate Revocation Lists (CRLs). In this paper, we present the first longitudinal characterization of the revocation statuses delivered by CRLs and OCSP servers from the time \revPAM{that the certificates expire}{of certificate expiration} to status disappearance. The analysis captures the status history of over 1 million revoked certificates, including 773K certificates mass-revoked by Let's Encrypt. Our characterization provides \revPAM{a novel snapshot and insights into}{a new perspective on} the Internet's revocation rates, quantifies how short-lived the revocation statuses are, 
\revPAM{captures biases and oddities in different sets of revoked certificates,
and highlights differences in how the statuses are handled by differ between CAs.}{highlights differences in revocation practices within and between different CAs, and captures biases and oddities in the handling of revoked certificates.}
Combined, the findings 
\revPAM{provide motivation for}{motivate} 
the development and adoption of \revPAM{transparent revocation logging schemes}{a revocation transparency standard}.


\end{abstract}
\maketitle      

\fancypagestyle{firststyle}
               {
                 \lhead{}
                 
                 \fancyfoot[C]{\small \vspace{10pt}
                   In: {\em Passive and Active Measurement Conference (PAM 2021)}.
                   }
               }
               \thispagestyle{firststyle}

\renewcommand{\headrulewidth}{0pt}
\renewcommand{\footrulewidth}{0pt}

\section{Introduction}
The modern Internet 
uses the Web Public-Key Infrastructure (WebPKI) as a foundation to establish trust between clients and servers.
In WebPKI, Certificate Authorities (CAs) issue signed X.509 certificates that verify the mapping between 
public keys and public distinguished names, such as domain names.
%

In certain cases (e.g., a private 
\revPAM{key being compromised}{key compromise}, 
\revPAM{a mapping no longer being valid}{owner's request}, or misissuance by a CA), certificates must be revoked; i.e., rendered invalid. To protect clients and servers from the use of revoked certificates, WebPKI supports \revPAM{a number of}{several} revocation protocols.  \cameraPAM{Traditionally, the revocation status}{Currently, revocation statuses} of \cameraPAM{a certificate}{most certificates} can be \revPAM{learned}{obtained} \cameraPAM{by querying CA-operated}{via} Online Certificate Status Protocol \cameraPAM{(OCSP)~\cite{RFCOCSP} servers}{(OCSP) servers~\cite{RFCOCSP},} \cameraPAM{or (depending on CA) by downloading a}{but some CAs continue to support the traditional} Certificate Revocation \cameraPAM{List (CRL)~\cite{RFCPKI}}{Lists (CRLs)~\cite{RFCPKI} as a complementary option}. However, \cameraPAM{in practice, these}{these} pull-based protocols raise many security, privacy, and performance issues. Therefore, many browser vendors do not utilize \cameraPAM{these}{the} protocols~\cite{revocations15}, but instead, \revPAM{push}{they push} a \revPAM{limited}{proprietary} set of \revPAM{(by them) centrally gathered revocations}{revocations} to the users~\cite{OneCRL,CRLSets}. \cameraPAM{\revPAM{Naturally, t}{T}}{Yet, t}hese push-based revocation \revPAM{updates must be limited in size, leaving}{mechanisms have their own limitations, which leave} secure certificate revocation an open problem~\cite{chuat2019sok}.

Furthermore, as of today, there does not \revPAM{exists}{exist} any standardized \revPAM{mechanisms}{mechanism} in place (similar \revPAM{to what}{to} Certificate Transparency (CT)~\cite{RFCCT,GOAC17,scheitle2018rise} \revPAM{provides for}{w.r.t.} certificate issuance) \revPAM{to collect}{to provide} an immutable history of all revocations and corresponding revocation reasons. Consequently, there is no ability to easily study and detect revocation-related misbehavior by CAs (e.g., advertisement of wrong, or contradictory revocation statuses).
While many novel WebPKI extensions, revocation protocols, architectures, and transparency schemes have been proposed to address this issue, none have been adopted so \revPAM{far}{far~\cite{chuat2019sok}.} 
Instead, \revPAM{as observed in the datasets presented here,}{we observe that the} information about revocations is sparse and \revPAM{revocation statuses often disappears soon after a certificate expires}{most revocation statuses disappear soon after certificate expiration}.


\cameraPAM{In this paper, 
we present a novel characterization 
\revPAM{in which we study}{study of} 
the revocation rates on the Internet, the post-expiry life of revocation statuses, \revPAM{how the status handling differ between CAs;}{the status-handling practices across CAs,} \revPAM{and make a case for the need of}{and make a case for} revocation transparency.}{In this paper, we make {\it a case for revocation transparency} by presenting a novel characterization study of the revocation rates on the Internet, the post-expiry life of revocation statuses, and the status-handling practices across CAs.}  First, we present a measurement methodology that allows us (i) to obtain \revPAM{close-to-all}{nearly all} revocations \revPAM{that have been done}{performed} for the set of certificates expiring \revPAM{at particular days}{during a time window}, and (ii) to track the certificate status (using both OCSP and CRL) of such sample sets over 100-day periods, starting at their respective expiration dates\footnote{Currently, CAs must maintain
\revPAM{a revocation status for a certificate only until it expires~\cite{CABR20}.}{revocation statuses only until certificate expiration~\cite{CABR20}.}}.

\begin{figure*}[t]
  \centering
\includegraphics[trim = 0mm 2mm 32mm 4mm, width=0.80\textwidth]{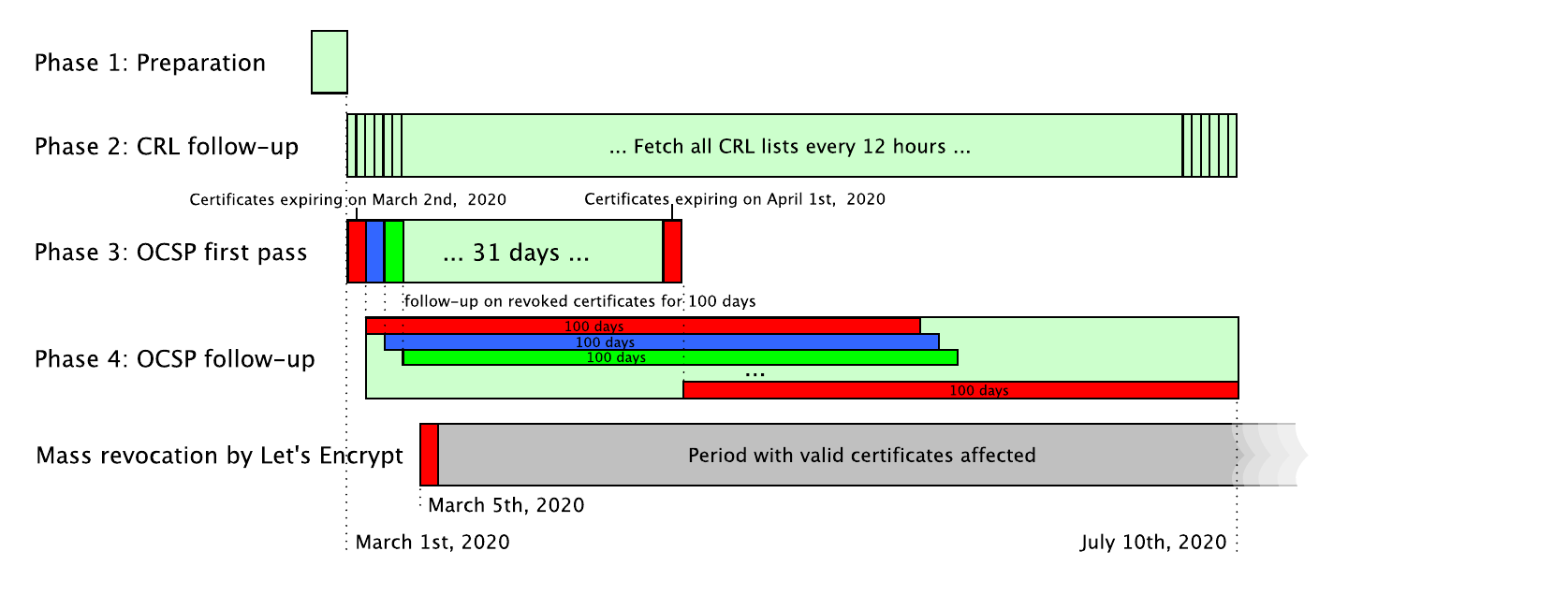}
  \caption{Timeline of the measurement.}
  \label{fig:phases}
\end{figure*}

Second, 
\revPAM{we use the methodology to track}{we track} 
all certificates 
\revPAM{in}{from}
the Censys dataset~\cite{censys15} that expired between Mar. 2, 2020, and Apr. 1, 2020, and that were valid with respect to Apple's, 
\revPAM{Microsoft's and}{Microsoft's, or} 
Mozilla's root 
\cameraPAM{\revPAM{stores.}{store.}}{stores.}
This time period (see Figure~\ref{fig:phases}) is particularly interesting since the measurement was done prior to and during the mass-revocation event in which Let's Encrypt (LE), the largest CA, \revPAM{revoked}{initially announced to revoke} over 3 million certificates~\cite{LE-datasets2020} \revPAM{}{due to a CAA-rechecking bug, but in the end, they revoked only 1.7M certificates~\cite{LE-Forum}}. 

Third, and most importantly, we characterize the \cameraPAM{revocation status handling}{revocation-status-handling} \revPAM{by different}{practices across} CAs, including \cameraPAM{how long they keep status information}{status lifetimes} beyond the expiration date \revPAM{(capturing the window to log the status of all revoked certificates) and}{and} \cameraPAM{how the handling differs}{handling differences} across CAs and certificate types.
\revPAM{In particular, we}{We} 
identify classes of behaviors, compare and contrast \revPAM{the handling}{practices} of different CAs, \revPAM{}{find} revocation biases among different sets of certificates, and look closer at some odd CA behaviors (e.g., certificates \revPAM{which was}{that} \revPAM{switched}{switch} back to a ``Good" status after \revPAM{having been listed}{being advertised} as ``Revoked").  Across our analysis, we observed highly heterogeneous behaviors among CAs and \revPAM{quickly disappearing}{quick disappearance of} revocation \revPAM{statues}{statuses}.  This highlights the lack of \revPAM{globalized standards and need for transparent revocation schemes}{a global revocation transparency standard} that \cameraPAM{\revPAM{can}{would}}{would otherwise} \cameraPAM{help}{help to} \revPAM{better identify}{identify} and improve odd revocation behaviors,
\revPAM{like what}{similarly to} 
\cameraPAM{\revPAM{CT has provided for the}{what CT provides the WebPKI issuance process}}{CT, with its effect on the issuance process}.
\revPAM{}{Finally, we share our dataset~\cite{KoCa21-dataset}.}

{\bf Outline:}
After a brief overview of revocation protocols (Section~\ref{sec:background}), we present our methodology (Section~\ref{sec:method}) and characterization results (Section~\ref{sec:results}).  Finally, related work (Section~\ref{sec:related}) and conclusions (Section~\ref{sec:conclusions}) are presented.

\section{Revocation protocols}\label{sec:background}

The two primary revocation protocols that CAs typically use 
are the following:
\begin{itemize}
\item {\bf Online Certificate Status Protocol (OCSP):}
Using OCSP, a client can request the status of a certificate by providing a serial number and the hashes of the issuer's name and key. The CA-Browser forum requires signed responses to be valid for at least \revPAM{every 8}{8} hours, and at most 10 days~\cite{CABR20}.  OCSP can be used in different ways.  For example, 
{\it OCSP stapling} allows 
\revPAM{OCSP statuses}{statuses} 
to be delivered by a web-server, and the {\it OCSP Must-staple} extension prevents a client from making OCSP requests on their own and enforces a hard-fail policy if the status was not delivered by the web-server. \revPAM{}{The Must-staple extension is not widely adopted yet~\cite{muststaple18}. Instead, most browsers typically accept a certificate if they are unable to obtain revocation information~\cite{revocations15}.}
\item {\bf Certificate Revocation List (CRL):}
CAs maintain signed \revPAM{list}{lists} with the serial numbers of revoked certificates, and optionally, corresponding invalidation dates and reason codes for the revocations. 
CRLs can also be augmented using several extensions (e.g., CRL number, Authority Key Identifier, etc.)~\cite{RFCPKI}.
CRLs are required to be reissued at least once every 7 days~\cite{CABR20}.
\end{itemize}

Due to \cameraPAM{previously mentioned}{the} security, privacy, and performance \cameraPAM{problems}{issues with OCSP and CRL}, many browser vendors have disabled the above pull-based revocation \cameraPAM{checks (OCSP and CRL), to}{protocols;} \cameraPAM{instead}{instead, they} periodically push limited sets of revocations to \cameraPAM{their}{the} clients (e.g., via software updates)~\cite{OneCRL,CRLSets}. \revPAM{However, 
this approach is limited by the revocation information gathered by the vendors and requires vendors to prioritize revocations (and the timing thereof) on behalf of the clients.}{However, this approach has some limitations; e.g., a delay introduced by scheduled updates, and a small coverage of all existing revocations.}

\revPAM{In contrast to the issuance process, where CT is prevalent, the WebKPI lacks revocation transparency.}{WebPKI lacks revocation transparency, and no mechanism similar to CT has been adopted yet.}
In fact, 
CAs are not required to maintain revocation statuses \revPAM{for their own}{for} certificates beyond their expiration date~\cite{CABR20}, and as we show \revPAM{here}{in this paper}, most of \revPAM{}{the} time, revocation statuses stop being advertised shortly after \revPAM{this date}{certificate expiration}. The lack of a transparent and immutable \revPAM{logging}{history} of revocations complicates 
keeping \revPAM{CAs (and browsers)}{CAs} accountable for their revocation \revPAM{handling}{mishandling}.

\section{Measurement methodology}\label{sec:method}

\revPAM{We conducted an active measurement campaign split over four phases (see Figure~\ref{fig:phases}), where the later three phases are overlapping.}{We conducted a four-phase measurement campaign (see Figure~\ref{fig:phases}).}

{\bf 1. Preparation:}
In the first phase, we collect all X.509 certificates (with their parent certificates) found in CT logs~\cite{RFCCT} and active scans that expire within a period starting from Mar. 2, 2020, to Apr. 1, 2020, using Censys~\cite{censys15}. For the analysis, we only select certificates that are valid with respect \revPAM{to (at least one of)}{to} Apple's, Microsoft's, or Mozilla's root stores~\cite{korzhitskii2020characterizing}. 
From these certificates, we extract all OCSP responder URLs (used in phases 3+4) and CRL URLs (used in phase 2).
For every remaining certificate, we then schedule an OCSP first pass (phase 3) 22 hours before  \revPAM{each certificate's individual, individual}{its} expiration\footnote{\revPAM{}{The interval of 22 hours (slightly less than 24 hours) was selected for performance reasons, after the initial evaluation of our measurement framework.}}, and for every observed CRL, we schedule periodic CRL requests (phase 2).

{\bf 2. CRL follow-up:}
During the second phase, we regularly (every 12 hours) fetch all CRL lists using the URLs extracted in the first phase. 

{\bf 3. OCSP first pass:}
In the third phase, 
we perform an OCSP status lookup for each certificate 22 hours before it expires. If a certificate is found to be revoked during its first pass, it gets scheduled for follow-up checks every 12 hours (phase 4). In the case of an OCSP timeout or an error, the first pass is retried every minute until a revocation status is obtained or the certificate is expired.

{\bf 4. OCSP follow-up:}
In the \revPAM{forth}{fourth} phase, 
the revocation status of every revoked expired certificate is fetched every 12 hours for 100 days (since the first pass of each individual certificate).
We separate OCSP responses into four types: ``Good", ``Revoked", ``Unauthorized", and ``Unknown".  The first two types (\cameraPAM{``Good", ``Revoked"}{``Good" and ``Revoked"}) are \cameraPAM{signed}{cryptographically-signed} responses that \cameraPAM{tell a client}{definitively specify}  \cameraPAM{about the}{the} \cameraPAM{definite status}{status} of a certificate. 
The third type (``Unauthorized") \cameraPAM{, is a plaintext unsigned}{is an unsigned plaintext} response.  The final category (``Unknown") contains signed ``Unknown" \cameraPAM{statuses, that some CAs deliver, and other responses, that are not signed}{statuses (that some CAs deliver) and other unsigned responses}.

{\bf External effects on the sampling rate:}
Between \cameraPAM{May 12 and May 19}{May 12, 2020, and May 19, 2020,} parallel processes \cameraPAM{run at our server caused the average OCSP inter-request periods to temporarily increase to an average response time between 12 and 21.7 hours on average (rather than 12)}{running at our server have temporarily increased the average OCSP inter-request time from 12 hours up to 21.7 hours}.  \revPAM{With the exception of}{Except for} this short \cameraPAM{time period}{period}, the average OCSP inter-request time was consistently 12 hours \cameraPAM{plus/minus}{$\pm$} a few minutes, up until \cameraPAM{June 21}{June 21, 2020}.  Between \cameraPAM{June 21}{June 21, 2020,} and the end of our measurement period \cameraPAM{(July 20)}{on July 20, 2020}, the average inter-request time was roughly 24 hours.  Neither of the periods with increased OCSP inter-request times took place during the first month after \cameraPAM{any of the certificates’ expiry time and}{the expiration date of any of the certificates; hence, the effects} do not impact our conclusions.



\section{Characterization results}\label{sec:results}

\subsection{High-level breakdown}

\begin{table}[b]
\caption{Summary \cameraPAM{of}{of the studied} certificates. }\label{tab:summary-dataset}
\centering
{\scriptsize
    \begin{tabular}{|l|r|r|r|r|}
    \hline
    \textbf{Certificates} & \textbf{Mass-revoked LE} & \textbf{Rest LE} & \textbf{Other CAs} & \textbf{All}\\\hline
     Non-revoked & \multicolumn{1}{c|}{--}      & 36,755,317 & 11,496,607 & 48,251,924\\\hline
     Revoked     & 773,128 & 129,552    & 174,712    &  1,077,390\\\hline
    Revocation rate  & 100\%   & 0.35\%     & 1.50\%     & 2.18\% \\\hline
    \end{tabular}}
\end{table}

In total, we collected OCSP status information for 49M certificates. Table~\ref{tab:summary-dataset} provides a breakdown based on whether a certificate was revoked or not, whether the certificate was issued by Let's Encrypt (76.3\% of the certificates) or a different CA (23.7\%), and whether a Let's Encrypt certificate was part of the above-mentioned mass-revocation event 
\revPAM{(1.57\%).\footnote{We used a dataset~\cite{LE-datasets2020} by Let's Encrypt to identify the mass-revoked certificates.}}{(1.57\%).} 
\revPAM{}{For us to consider a certificate mass-revoked it needed to be (i) on the list of 3M certificates that Let's Encrypt publicized for the event~\cite{LE-datasets2020} and (ii) to be revoked at the time it expired. \cameraPAM{In addition to the mass-revoked certificates reported here, we also observed 297K Let's Encrypt certificates that had not been revoked}{We also found that 297K certificates from the list, with expiration dates falling on our first pass period, have never been revoked}.}

The timing of the mass-revocation event is particularly interesting since it provides a \revPAM{concrete use-case}{concrete} example of the impact \revPAM{that this event had}{that such events can have} on the revocation rate and 
\revPAM{whether the revocation statuses of these certificates were removed faster/slower than other certificates}{the lifetime of revocation statuses}. 
\revPAM{}{Finally, we note that the certificates affected by recent mass-revocation events have been disclosed through website postings \cameraPAM{using}{of} arbitrarily formatted datasets~\cite{LE-datasets2020,massDig2,massDig1}.}

While the \cameraPAM{revocation rate for}{non-mass-revocation-rate of} \cameraPAM{the other Let's Encrypt certificates were}{Let's Encrypt was} much smaller than for the other CAs (0.35\% vs 1.50\%), the mass-revocation event increased Let's Encrypt's revocation rate for this period \cameraPAM{to}{up to} 2.40\%. \cameraPAM{This big jump is}{The effect is} perhaps most noticeable when looking at the number of revoked certificates per day, based on their day of expiry, as shown in Figure~\ref{fig:revoked-per-day}.  Here, \cameraPAM{we}{starting from Mar. 5, 2020, we can} see \cameraPAM{a big jump on Mar. 6 (starting on Mar. 5) caused by}{the impact of the} certificates associated with the mass-revocation event. The other two classes of revocations remained relatively stable throughout the measurement period.

\begin{figure}[t]
  \centering
  \includegraphics[trim = 7mm 20mm 3mm 0mm, clip, width=0.5\textwidth]{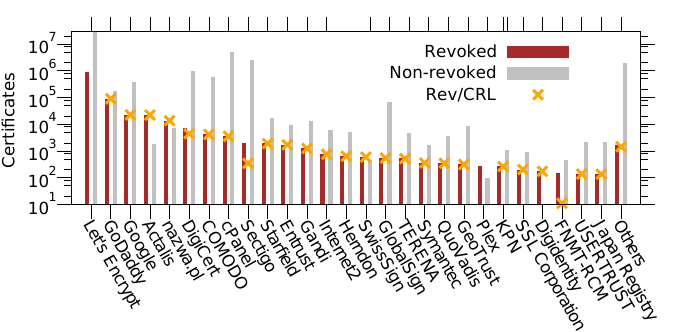}
  \caption{Per-CA breakdown of the number of revoked and non-revoked certificates in the dataset. Revoked certificates found in CRLs are shown with \textcolor{orange}{$\times$}. \revPAM{}{CAs are ordered by the number of revoked certificates, in descending order from left to right. In the following figures, the order is preserved.}}
  \label{fig:revoked-per-CA}

\end{figure}

\cameraPAM{There are also big}{We found large} variations in the revocation rates of different CAs.
Figure~\ref{fig:revoked-per-CA} shows the number of revoked and non-revoked certificates, 
broken down per CA.  
Moreover, we mark the number of revoked certificates \cameraPAM{for which CRL data also were provided}{listed in the CRLs, in addition to OCSP servers} \cameraPAM{discussed in}{(discussed in} Section~\ref{sec:CRL}).
Here, we show all CAs with at least 100 revoked certificates in our dataset, ranked from the one with the \cameraPAM{most (Let's Encrypt)}{most revocations} to the one with the least.  We also \cameraPAM{include an}{include the} ``other" category that combines the results for all other CAs.  While most CAs have much fewer revoked certificates than non-revoked certificates, there are notable exceptions.
Five CAs even had more revoked than non-revoked certificates: Actalis (92.5\%), nazwa.pl (66.4\%), SwissSign (59.9\%), Plex (73.7\%), Digidentify (100\%).  Among the most popular CAs, GoDaddy stands out with 34.5\% 
of its certificates being revoked before expiration.


\subsection{Revocation status changes}

\begin{figure}[b]
  \centering
  \subfigure[CDFs]{
    \includegraphics[trim = 13mm 8mm 0mm 5mm, clip, width=0.5\textwidth]{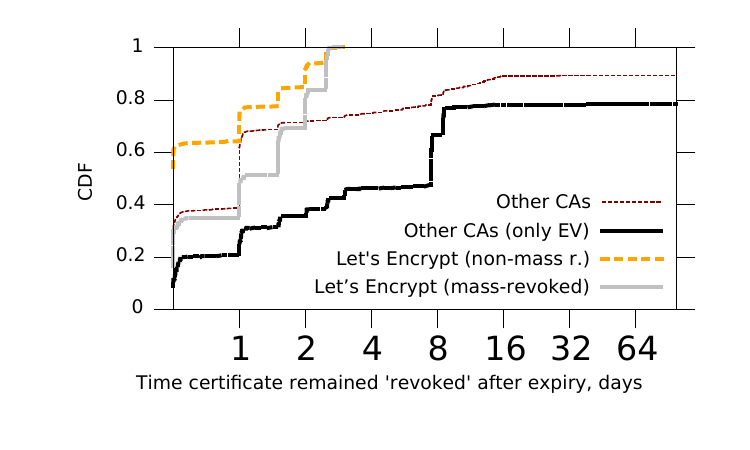}}
  \hspace{3pt}
  \subfigure[Per-CA breakdown]{
    \includegraphics[trim = 50mm 940mm 0mm 20mm, clip,  width=0.5\textwidth]{figures/revoked-status2-perCA-v02-legendfix2.pdf}
    }
   
 \vspace{-10pt}
  \caption{Time that the revoked certificates remained revoked after \revPAM{they expire}{expiration}.}
  \label{fig:time-revoked}
\end{figure}

The revocation statuses provided by
OCSP servers often change from ``Revoked'' to some other status soon after certificate expiry.
Figure~\ref{fig:time-revoked} shows the time that the status remained ``Revoked" after the revoked certificates had expired. Here, we filter out any temporary OCSP responses (e.g., unauthorized, unknown) and timeouts whenever we obtained at least one more ``Revoked" response.
 
 \begin{figure}[t]
  \centering
    \includegraphics[trim = 0mm 1226mm 0mm 6mm, width=0.5\textwidth]{figures/expirations-v01-legendfix3-bw.pdf}
  \caption{\revPAM{Number revoked certificates per day of expiry.}{Revoked certificates per expiration date.}}
  \label{fig:revoked-per-day}
\end{figure}
 
{\bf Quickly disappearing revocation statuses:}
Figure~\ref{fig:time-revoked}(a) shows the empirical Cumulative Distribution Functions (CDFs) for four classes of revoked certificates: 
\revPAM{2$\times$}{2 for} 
Let's Encrypt certificates 
\revPAM{(based on whether they belong to the mass-revocation event or not)}{(mass-revoked and non-mass-revoked)}
and 
\revPAM{2$\times$}{2 for} 
certificates by other CAs 
\revPAM{(based on whether it is an Extended Validation (EV) certificate or not).}{(with and without Extended Validation (EV)).}
\revPAM{While the mass-revoked certificates (gray) had 
somewhat longer
status change times than the other Let's Encrypt certificates (orange), 
the ``Revoked" statuses were always changed within 3 days of expiry for certificates issued by Let's Encrypt.}{All certificates by Let's Encrypt changed status within 3 days of expiration.
Their mass-revoked certificates had longer status change times than the non-mass-revoked certificates.}
\revPAM{The big volume of mass-revoked certificates around this time highlights the short time window to capture the revocation of certificates revoked close to expiry.  The lack of a standardized revocation transparency method, was also highlighted by Let's Encrypt announcing their mass revocation event in arbitrary way; in this cases by publishing datasets of the revoked certificates on their website~\cite{LE-datasets2020}.
The}{The} 
CDFs \cameraPAM{for the other categories}{for the other CAs} are relatively flat from about two weeks to 100 days.  (Note the logarithmic y-axis.) On an encouraging note, the \cameraPAM{revocation}{certificate} class with the most long-lived revocation statuses \cameraPAM{were found among}{is} Extended Validation (EV) certificates.  \cameraPAM{This is also the}{This} \revPAM{type}{class} of \cameraPAM{certificates that}{certificates} \cameraPAM{typically should}{should typically} endure the most scrutiny.
%

{\bf Some CAs keep the \revPAM{state for}{state} longer:}
Figure~\ref{fig:time-revoked}(b) shows the fraction of the certificates issued by different CAs that maintained the revoked status for at least 1 week or 30 days.  While \revPAM{also many of the other}{many} CAs maintained ``Revoked" state for very short time periods after certificate expiry (see Figures~\ref{fig:time-revoked}(a) and \ref{fig:time-revoked}(b)), most of the CAs that did keep the ``Revoked" state beyond a week also kept this \cameraPAM{state for}{state} 
\revPAM{more than a month (e.g., CAs with both bars in Figure~\ref{fig:time-revoked}(b) close to 100\%).}{\cameraPAM{more than}{beyond} 30 days.}


{\bf Status response overview:}
For the revoked certificates,
we performed
more than 207M OCSP status requests.

All certificates started as ``Revoked'' and most eventually changed to an unauthorized response (100\% of Let's Encrypt certificates and 76.43\% of other CAs' certificates).  While we only had timeouts for 0.04\% of the status requests, the differences between the number of affected certificates were substantial between CAs: only 0.07\% of the Let's Encrypt certificates had at least one timeout, compared to 13.98\% of the other CAs' certificates.  
\revPAM{This non-negligible fraction demonstrates a weakness with browser policies that fall back to trusting certificates when the OCSP servers do not respond.}{These fractions are non-negligible, since most browsers soft-fail on an OCSP timeout and continue to establish a potentially-insecure connection.} 
\revPAM{Perhaps most surprisingly, and concerningly, we observed}{A concerning observation is that}
589 certificates \revPAM{}{issued by 13 CAs} (0.34\% in the other CA category) \revPAM{for which the CA 
went back to responding that the certificate was}{switched from ``Revoked'' status to} ``Good'' (66K responses in total).  

\begin{figure}[t]
  \centering
    \includegraphics[trim = 70mm 1060mm 30mm 0mm, width=0.5\textwidth]{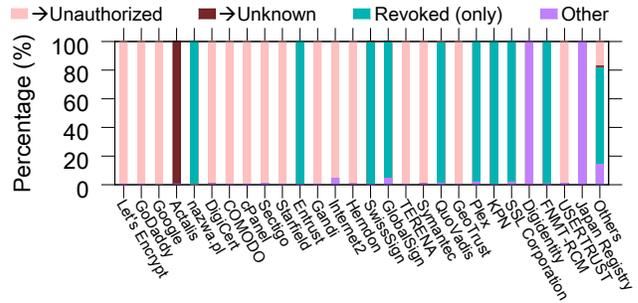}
  \vspace{-5pt}
  \caption{Dominating status change behavior of CAs.}
  \label{fig:status-perCA}
  \vspace{-5pt}
\end{figure}

{\bf Most frequent behaviors:}
\revPAM{}{Usually, public certification practice statements of CAs 
guarantee revocation status preservation for non-expired certificates, 
but do not specify the CAs' actions after that~\cite{ISRG20,GoDady-policy-20,Google-policy-21}.}
We next 
\revPAM{look closer}{look} 
at the most frequent CA behaviors. For this analysis, we filtered out temporary status changes whenever we observed the original state again. With this filtering, we observed the following \cameraPAM{case-by-case distributions among the dominating}{dominating} behaviors.
\begin{itemize}
\item Let's Encrypt almost always transition 
\revPAM{}{statuses}
from ``Revoked'' to ``Unauthorized''.
This \revPAM{behaviors}{behavior} was observed for 772K (99.86\%) of the mass-revoked certificates and 129K (99.88\%) of the 
other
\revPAM{Let's Encrypt}{certificates revoked} \revPAM{}{by Let's Encrypt}.
\revPAM{}{A possible explanation for this behavior is that they respond with code ``Unauthorized" as soon as the status record has been removed~\cite{DeHu07}.  Let's Encrypt's current certification practice statement only guarantees that ``OCSP responses will be made available for all unexpired certificates"~\cite{ISRG20}.}
\item Among the other CAs, we observed three dominating behaviors: 133K (76.28\%) cases where the CA simply transitioned to ``Unauthorized'' (like Let's Encrypt), 21K (12.49\%) cases where the status always changed to ``Unknown'', and 19K (10.68\%) cases where the ``Revoked'' status remained for the duration of our measurement period.
\end{itemize}
Figure~\ref{fig:status-perCA} breaks down the use of the dominating status change behaviors employed by the different CAs.  In addition to the three behaviors mentioned above, we \cameraPAM{include an}{include the} ``other'' behavior category.
Most CAs \revPAM{has one}{have a} dominating behavior that \revPAM{it}{they} employ\revPAM{s}{} for almost all \revPAM{its}{of their} certificates: 
15 (out of 26) CAs almost always switch from ``Revoked'' to ``Unauthorized'',
9 (out of 26) CAs almost always keep the ``Revoked'' status for the full 100 day period,
Actalis \revPAM{almost always}{mainly} \revPAM{switches respond with status}{switch certificates from ``Revoked" status to} ``Unknown" (except for 91 cases, when the statuses \revPAM{switches}{were switched} to ``Good", \revPAM{after first temporarily responded with}{following the intermediate} ``Unknown" status),
Digidentify (\revPAM{who revoked all their certificates}{who revoke all certificates}) always start\revPAM{s}{} to timeout, and Japan Registry always switches statuses to ``Good''.
As expected, the ``other CA" category (not explicitly listed), contains a mix of behaviors.  These results \revPAM{highlight that there is no common agreed branch standard for how to maintain revocation state after expiry}{demonstrate the lack of a standard practice w.r.t. revocation statuses after certificate expiration}.
\revPAM{}{We have also observed some small differences in the weekly status-change patterns between CAs; however, compared \revPAM{to}{to the} differences \revPAM{in the}{in} issuance timing, these differences are very small.  See Appendix~A.}


{\bf Special cases with the ``Good'' status:}
\revPAM{Out of the 589}{589} \revPAM{}{revoked} certificates \revPAM{for which the OCSP servers switched to responding with}{switched to} status \revPAM{``Good'', the majority maintained this status for a significant time period.}{``Good''.} 
\revPAM{In 349 cases the servers kept the ``Good" status for the remaining duration of our measurement campaign after having changed to ``Good" directly from ``Revoked", and in 91 casesit kept the ``Good" status for the remaining duration 
after having changed to ``Good" via a few instances reporting status ``Unknown".}{In almost all cases the servers kept the ``Good" status until the end of the measurement period.
In 349 of these cases, the status changed directly from ``Revoked" to ``Good" and in 91 cases an intermediate ``Unknown" status was observed.}
All these cases provide strong motivation for transparent long-term recording of revocation information.

\cameraPAM{At this time, we}{We} note \revPAM{that that}{that} Let's Encrypt and most of the other big CAs
did not have any cases with the above strange behavior.
Of the CAs with at least 100 revocations, only the following CAs had \revPAM{}{such} cases: GoDaddy (117 cases), Actalis (91), Starfield (9), Entrust (5), and Japan Registry (135).
Other CAs (not listed in our figures) with many cases include:
``National Institute of Informatics'' (91), ``SECOM Trust Systems'' (70), ``ACCV'' \revPAM{54}{(54)}.  
(The rest of the non-listed CAs 
had five or fewer revoked certificates changing to status ``Good''.)
Finally, 
a few certificates in this category stood
out more than the others.  
For example, the list included three EV certificates: one by Entrust for ``JPMorgan Chase and Co'' (``Revoked" $\rightarrow$ ``Good" $\rightarrow$ ``Revoked"),
one by GoDaddy for ``Delmarva Broadcasting Company'' (``Revoked" $\rightarrow$ ``Unauthorized" $\rightarrow$ ``Good"),
and one by Actalis for ``Pratiche.it'' (``Revoked" $\rightarrow$ ``Unknown" $\rightarrow$ ``Good").
Otherwise, all the certificates in this class 
\revPAM{used RSA: 9 with key length 1024, 579 with 2048, and 1 with 4096.}{include RSA keys with the following key lengths: 1024 (9), 2048 (579), and 4096 (1).} 
Furthermore, only 123 (out of 589) had \revPAM{SCTs}{Signed Certificate Timestamps (SCTs) embedded}.
\surveyPAM{}{We contacted all CAs with the above behavior.  A summary of the responses is provided in Appendix B.}



\subsection{Biases in the revocation sets}

\begin{figure}[t]
  \centering
  
  \subfigure[CDFs of certificate validity]{ 
   \includegraphics[trim = 12mm 15mm 0mm 5mm, clip,width=0.5\textwidth]{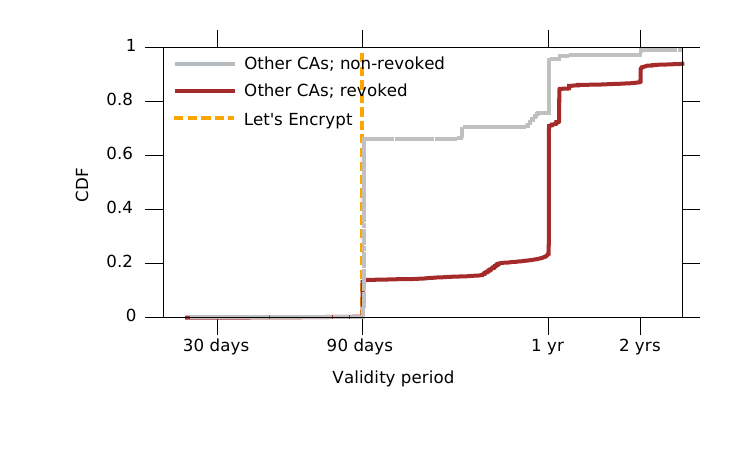} }
  \hspace{-2pt} \vspace{10pt}
  \subfigure[Revoked certificates \revPAM{, per CA}{}]{
    \hspace{3pt}\includegraphics[trim = 100mm 800mm 0mm 120mm,  width=0.5\textwidth]{figures/valid-perCA-v01-legendfix4.pdf}}
  
  \subfigure[Non-revoked certificates \revPAM{, per CA}{}]{
    \includegraphics[trim = 100mm 1000mm 0mm 50mm, width=0.5\textwidth]{figures/valid-perCA-nonRev-v01-legendfix4.pdf}}
  \caption{Validity periods for different categories of revoked and non-revoked certificates.}
  \label{fig:validity-periods}
  \vspace{-12pt}
\end{figure}

{\bf Validity period:}  We have found that the revoked
certificates typically have longer validity 
\revPAM{periods. This is illustrated in Figure~\ref{fig:validity-periods}. First,}{periods.}
Figure~\ref{fig:validity-periods}(a) shows CDFs of the validity periods for both revoked and non-revoked certificates for all \revPAM{other CAs}{CAs other} than Let's Encrypt. (Since Let's Encrypt always use a 90-day validity period, we kept these certificates separately.) 
Here, we note a clear shift between the two curves.

Figures~\ref{fig:validity-periods}(b) and~\ref{fig:validity-periods}(c) provide a similar comparison of the (b) revoked and (c) non-revoked certificates on a per-CA basis.  Here, we plot the fraction of certificates with validity periods longer than 89 days, 90 days, 1 year (365 days), and 2 years (720 days), respectively. These choices are based on the observation that many CAs use validity periods of either 90 days or 398 days (e.g., steps in the CDFs in Figure~\ref{fig:validity-periods}(a)).
For almost all CAs, the fraction of certificates with long validity periods is larger among the revoked certificates (Figure~\ref{fig:validity-periods}(b)) than among the corresponding CA's non-revoked certificates (Figure~\ref{fig:validity-periods}(c)).
This is in part an effect 
\revPAM{of CAs being pushed (by browser vendors)
to use shorter certificate validity periods~\cite{CABR20,apple398,google398,mozilla398}}{of CA/Browser Forum conventions~\cite{CABR20} and decisions by individual browsers~\cite{apple398,google398,mozilla398} forcing CAs to use shorter certificate validity periods.}
\revPAM{and older certificates having had more time to (in some way) become compromised.}{Another reason is that older certificates have had more time to become compromised.}
\revPAM{However, although not explicitly supported here, it}{It} 
could also be an indication that CAs apply increasingly \revPAM{better}{stricter} security policies (e.g., \revPAM{as pushed by initiatives such as}{to comply with} CT~\cite{RFCCT}).

{\bf Public key types:}
\revPAM{}{The modern WebPKI relies 
on EC (Elliptic Curve)~\cite{koblitz1987elliptic} and RSA (Rivest--Shamir--Adleman)~\cite{rivest1978method} public-key cryptography.}
\revPAM{}{Here, we compare the use of different key types and key lengths.} 
While RSA 2048 is the dominating public key among both 
revoked (90.44\%) and non-revoked (80.81\%) certificates, there are significant differences in the revocation rates of certificates including different key types. For example, certificates with RSA 3072 (4.55\% revocation rate), 
EC 521 (80.49\%) and RSA with key lengths other than the three most common lengths
(6.67\%) all have revocation rates well above average. In contrast, EC 256 (0.14\%), EC 384 (0.62\%) and RSA 4096 (1.48\%) all have revocation rates below average.  These differences are also present when looking at 
certificates of Let's Encrypt and other CAs separately.

{\bf SCT and EV usage:}
To measure the CT compliance we looked at the use of Signed Certificate Timestamps (SCTs). While all certificates issued by Let's Encrypt have \revPAM{SCTs}{embedded SCTs}, \revPAM{some of the other CAs}{other CAs} do not always \revPAM{include SCTs}{embed the timestamps}.  Furthermore, among the certificates issued by other CAs, the fraction of certificates that do not contain SCTs was much greater among the revoked 
(10.04\%) than 
non-revoked certificates (1.91\%).  
\revPAM{Part of the reason for these differences are that older certificates (among which we see higher revocation rates) are less likely to be CT compliant, but may also capture that some CAs replace old certificates that are not CT compliant with newer certificates.}{In addition to having longer validity periods, some of the older non-expired certificates lack embedded SCTs.
Owners and issuers of these certificates may be replacing them with 
certificates that better meet recent browser requirements~\cite{apple398,ChromeFullCT}.}
We have also observed significantly higher revocation rates among EV certificates. For example, 2K (10.77\%) out of the 18K observed EV certificates \revPAM{are}{were} revoked. 
\revPAM{Furthermore, out of the certificates revoked by CAs other than Let's Encrypt, 1.08\% are EV certificates, but only 0.14\% of the non-revoked certificates.}{Furthermore, for CAs other than Let's Encrypt, 1.08\% of the revoked certificates are EV certificates and 0.14\% of the non-revoked certificates are EV certificates.}
\revPAM{Table~\ref{tab:summary-certs} summarize these numbers.
}{}

\subsection{CRL-based analysis}\label{sec:CRL}

For the 2K CRL URLs extracted from the certificates of interest, we collected 644K CRL snapshots.
\revPAM{In total, these snapshots together}{Combined, these snapshots} 
included CRL entries for 170K (15.8\%) of the revoked certificates \revPAM{}{found using OCSP}. 
\revPAM{The relatively lower CRL usage is due to Let's Encrypt not using CRLs.}{Let's Encrypt's decision not to implement CRL contributes to the small fraction.} 
Here, we focus on the certificates 
\revPAM{for which we found}{with} 
at least one CRL entry \revPAM{}{and one OCSP ``Revoked" status}.

\begin{figure}[b]
  \centering
  
  \subfigure[Normalized revocation time during certificate lifetime]{
    \hspace{-10pt}\includegraphics[trim = 14mm 20mm 0mm 6mm, clip, width=0.5\textwidth]{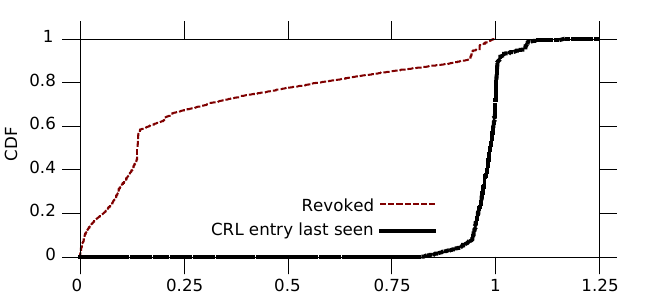}}
   \subfigure[Entries per CRL measurement]{
    \hspace{-10pt} \includegraphics[trim = 16mm 20mm 15mm 10mm, clip, width=0.5\textwidth]{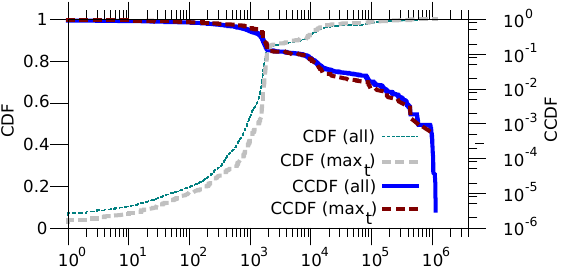}}
   \vspace{-12pt}
  \caption{\revPAM{Example distributions}{Distributions} for measured CRLs.}
  \label{fig:cdf-timing}
  \vspace{-12pt}
\end{figure}

{\bf Timing analysis:}
\revPAM{Revocation}{On average, revocation} \revPAM{information disappear}{statuses disappear} even faster from CRL lists than from OCSP \revPAM{servers}{responders}.  
For example, only in 26.5\% of the cases did we observe the revocation status in the CRLs after the expiration date of the certificates, and only for 2.9\% did we observe the status being preserved longer than a week after expiration.  
This may be an attempt to reduce the size of the CRLs.  
However, since the majority of the revocations happen early in the lifetime of the certificates (e.g., the median normalized lifetime is 13.8\%) there is still a significant time period over which certificates are included in the CRLs.  This is illustrated in Figure~\ref{fig:cdf-timing}(a), which shows the normalized timing of revocations and when the CRL entries are last observed in our dataset. Here, all values are normalized relative to the total intended validity period (i.e., ``NotBefore" and ``NotAfter" corresponds to the values 0 and 1, respectively).  
\revPAM{As obvious from Little's law, which implies that}{As implied by Little's law,} 
the average size of a CRL 
\revPAM{(e.g., in terms of number of entries per CRL)}{(e.g., measured as entries per CRL)}
is equal to the average time that the entries 
\revPAM{of the CRL remain}{remain} 
in the CRL (e.g., measured in days) times the average rate that certificates are being added to the CRL 
\revPAM{(e.g., measured in revocations per day),}{(e.g., revocations per day),} 
CRL sizes therefore easily become very large.  
Indeed, the average CRL size was 7K entries and the largest CRL 
contained 1.1M entries at its peak.  Figure~\ref{fig:cdf-timing}(b) shows CDFs and CCDFs for both 
individual measurements (all) and when using the 
\revPAM{peak size (max$_t$) observed for each CRL.}{observed peak size (max$_t$).}  
We also observed some CRLs that did not appear to delete entries and 
roughly 0.94\% of the certificates remained in the CRLs for the full duration of our measurement.

\begin{figure}[t]
\includegraphics[trim = 10mm 28mm 0mm 0mm, width=0.5\textwidth]{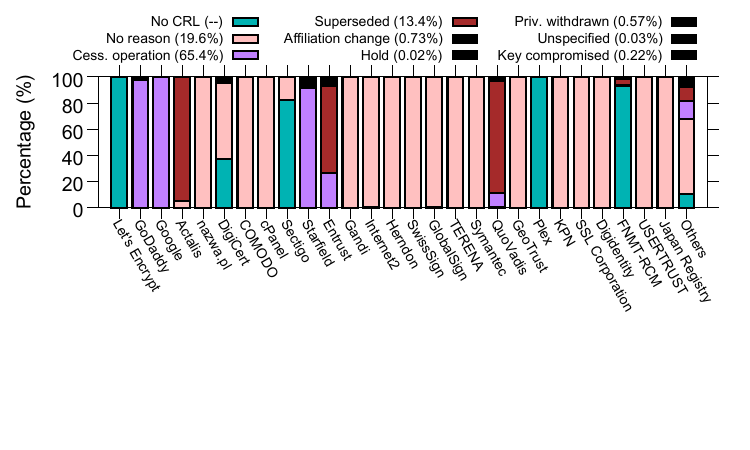}
  \caption{CRL reasons for revocation.}
  \label{fig:reason-perCA}
  \vspace{-12pt}
\end{figure}

{\bf \revPAM{Reason for revocation}{Revocation reasons}:}
Figure~\ref{fig:reason-perCA} breaks down on a per-CA basis the percentage of certificates for which (i) we did not find any CRL entry, (ii) we found CRL entries without a revocation reason, or (iii) we found a revocation reason for. The overall percentages for particular reasons (over all certificates with CRL entries) are provided in the figure's key.  The four dominating CRL behaviors that we observed were: (i) some CAs did not use CRLs (Let's Encrypt, Plex) or only used it to a limited degree (e.g., Sectigo, FNMT-RCM), (ii) 17 CAs used CRLs for the majority of their revocations but did not provide any revocation reason, (iii) three CAs almost always used ``Cessation Of Operation" as revocation reason (GoDaddy, Google, Starfield), and (iv) three CAs almost always specified ``Superseded" as the revocation reason.

\revPAM{As the majority of revoked certificates are not included in CRLs, and, out of those that are, 19.6\% give no revocation reason, the results also demonstrate the complexity to discern patterns across the whole certificate landscape (e.g., to identity clusters of compromised keys).}{Overall, 
\surveyPAM{the majority of}{most} 
revoked certificates were not included in 
\surveyPAM{CRLs. At the same time,}{CRLs and}
19.6\% of CRL entries contained no revocation reason.}
\revPAM{The big differences in how CAs behave captures that there is no agreed-upon standard for revocation bookkeeping that all CAs adhere to.}{Our results show that the practices of CAs are highly heterogeneous and revocation statuses are not persistent; thus, we argue that the Internet would benefit from a revocation transparency standard.}

\section{Related work}\label{sec:related}

\revPAM{Previously, a number of studies were performed to measure}{A number of studies have measured the}
revocation rates on the  
\revPAM{Internet; e.g., in~\cite{revocations15} the authors}{Internet. Liu et al.~\cite{revocations15}} 
performed several IPv4 HTTPS scans and found that a large fraction of served certificates was revoked (8\%), while CRLSets~\cite{CRLSets} by Google was only covering 0.35\% of all revocations. 
\revPAM{In~\cite{muststaple18}, the authors}{Chung et al.~\cite{muststaple18}} 
evaluated the performance of OCSP responders by sending OCSP requests from 
\revPAM{several geographically spread locations.}{geographically separated locations.}
They concluded that OCSP responders were not sufficiently reliable to support {\it OCSP Must-staple} extension. 
\revPAM{In an older measurement~\cite{zhu2016measuring}, the authors}{Zhu et al.~\cite{zhu2016measuring}} 
\revPAM{considered}{found} 
OCSP latency to be ``quite good", and showed that 94\% of OCSP responses are served using CDNs. Moreover, only 0.3\% of certificates were found to be revoked at 
\revPAM{the time.}{that time (2015).} 
\revPAM{In~\cite{smith2020let}, the authors}{Smith et al.~\cite{smith2020let}} 
propose an efficient scheme to disseminate revocations. In the process, they measured revocation rates and found that in the absence of a mass-revocation event, the revocation rate on the Internet was 
\revPAM{1.29\%, which is similar to the value}{1.29\%.  This is similar to} 
what we observed. The above works perform OCSP status checks before certificate expiration, while we check the certificates the day before their expiration and onward. Revocation effectiveness at the code-signing PKI was measured in~\cite{windowsRev}, and a number of security problems related to revocations were 
\revPAM{found in the work.}{identified.}
A recent survey 
and a
comprehensive framework for comparison of implemented and proposed revocation/delegation schemes \revPAM{is}{are} provided in~\cite{chuat2019sok}. 

{\bf Other community efforts and data sources:}
The CA/ Browser forum specifies some requirements that motivated our measurement design, including the requirement that ``revocation entries on a CRL or OCSP Response MUST NOT be removed until after the Expiry Date of the revoked Certificate"~\cite{CABR20}.  We used the Censys search engine, backed by Internet-wide scanning~\cite{censys15}, to obtain all certificates for our study.  Some other online services also provide revocation statuses.  For example, crt.sh~\cite{crt.sh} 
\extraPAM{selectively obtains CRL statuses for certificates; but, updates are not regular.}{fetches
known CRLs regularly,
and performs OCSP requests on-demand.}
\extraPAM{Until late}{Until} 
Aug. 2020, Internet Storm Center~\cite{isc} was regularly fetching several CRLs; however, they did not monitor all CRLs present in our dataset and did not capture the mass-revocation by Let's Encrypt. 

\section{Conclusion}\label{sec:conclusions}

In this paper, we have presented the first characterization of the revocation status responses provided by \revPAM{OCSP servers and CRLs}{OCSP and CRL responders} \revPAM{as seen from}{from} the time \revPAM{certificates expires}{of certificate expiration} and beyond.  \revPAM{Using our methodology and dataset}{We described a measurement methodology}, \revPAM{provided a novel snapshot and insights into}{which allowed us to look at} the revocation rates on the Internet \revPAM{}{from a new perspective}; we quantified how short-lived the revocation statuses are, and highlighted differences \revPAM{in how the statuses are handled by}{in status handling practices of} different CAs. We found that most CAs remove \revPAM{the revocation status of revoked certificates}{revocation statuses} very soon after \revPAM{expiration}{certificate expiration}. \revPAM{In the case of CRLs, some}{Some} CAs do not provide CRL entries for all revoked certificates and/or remove entries from the CRLs before \revPAM{the expiry of the certificates}{certificate expiration}. \revPAM{The many CA-based}{The CA-dependent} differences highlighted throughout the paper (e.g., \revPAM{how and for how long they maintain status information}{revocation status lifetimes}, \revPAM{their use of CRL revocation reasons}{usage of reason codes}, and \revPAM{strange behaviors in which some CAs switched back to give certificates status ``Good'' although they were ``Revoked'' at the time of expiry}{abnormal behavior of switching certificates from ``Revoked'' to ``Good'' status}) capture a highly heterogeneous landscape \revPAM{without standards for transparent bookkeeping of revocations}{that lacks a revocation transparency standard}. \revPAM{Like how CT has increased the transparency of issuance, these findings provide a strong case for the adoption of transparent revocation logging.  Finally, we}{Finally, we \cameraPAM{motivate}{argue for} the deployment of \cameraPAM{such}{revocation transparency} and} demonstrate the global impact of the mass revocation \cameraPAM{event that}{event, which} took place during our measurement campaign. \revPAM{In the paper}{We} compared the characteristics of the mass-revoked certificates with the characteristics of other revoked and non-revoked certificates issued by Let's Encrypt and the rest of the CAs, \revPAM{but only find limited}{and found a limited number of} biases, \revPAM{. Instead,}{e.g.,} the biggest differences in the revocation rates \revPAM{seem to be  related to}{depend on} \revPAM{whether certificates are issued by a particular}{the origin} CA, \revPAM{what key type is being used}{key type}, \revPAM{whether the certificate is an EV certificate or not}{EV policy}, and \revPAM{whether it is complying with CT policies}{presence of embedded SCTs}.

\section*{Acknowledgment}
This work was 
supported by the Wallenberg~AI, Autonomous Systems and Software Program (WASP) funded by the Knut and Alice Wallenberg Foundation.

\thispagestyle{firststyle}

\balance
\bibliographystyle{splncs04}
\bibliography{references}





\newpage
\thispagestyle{firststyle}

\balance

\begin{figure}[H]
  \centering
  \subfigure[End of revocation status]{
        \includegraphics[trim = 4mm 22mm 4mm 10mm, width=0.486\textwidth]{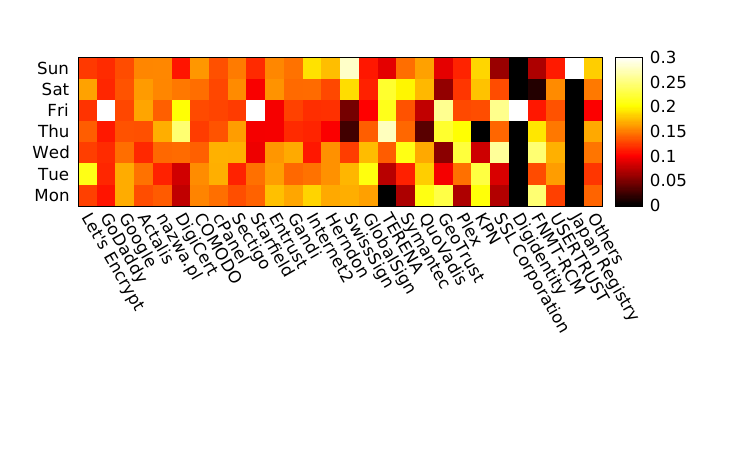}}
  \subfigure[Start of certificate validity period]{
    \includegraphics[trim = 4mm 21mm 4mm 4mm, width=0.486\textwidth]{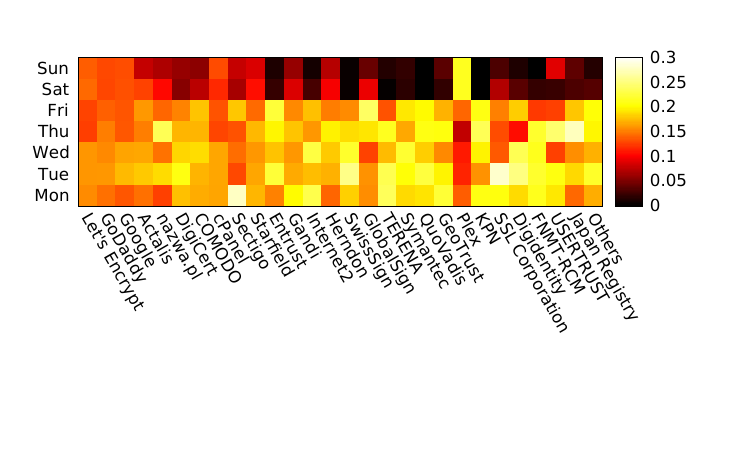}}
  \caption{Weekly distribution of certificate-validity-start day for the revoked certificates and last-status-change day (from ``Revoked'' to something else).}
  \label{fig:per-day-heat}

\end{figure}

\begin{figure}[b]
  \centering
  \subfigure[Expiry day]{
    \includegraphics[trim = 5mm 22mm 8mm 10mm, width=0.466\textwidth]{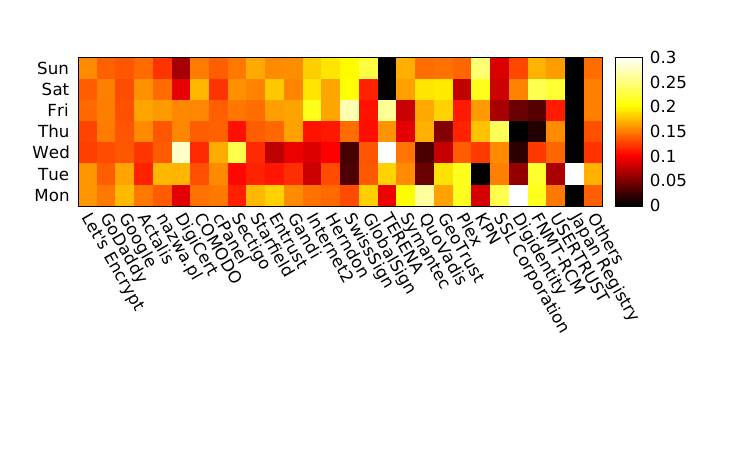}}
  \subfigure[Expiry time]{
    \includegraphics[trim = 19mm 23mm 16mm 4mm, width=0.486\textwidth]{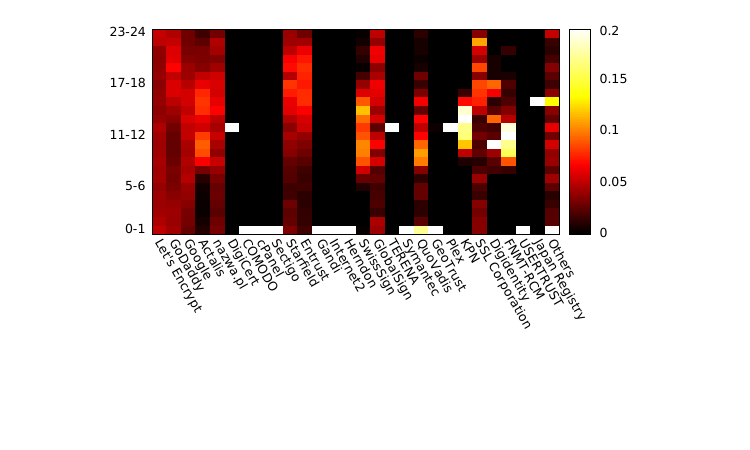}}
  \caption{Expiration time of revoked certificates.}
  \label{fig:expiry-heat}

\end{figure}

\section*{Appendix A. Extra on CA behavior}\label{sec:other-CA-comparisions}

We have already seen that different CAs have different practices for handling of revocation statuses. Here we present the obtained day-of-week distributions that capture {\it when} CAs change the ``Revoked'' status to something else (Figure~\ref{fig:per-day-heat}(a)); compare this to the distribution of the first certificate validity day (Figure~\ref{fig:per-day-heat}(b)). We note the weak weekly patterns.
While more than half of the CAs issue significantly fewer certificates with start dates during weekends (dark areas for Sat/Sun in Figure~\ref{fig:per-day-heat}(b)), we did not observe such weekly patterns for the revocation status changes. Instead, only a few CAs have spikes of revocation status changes on a certain day (white squares in Figure~\ref{fig:per-day-heat}(a)). For example, Starfield, GoDaddy (part of Starfield), and Digidentify update most of their statuses on Friday, and Japanese Registry on Sunday (Monday Japanese time). 
The distributions suggest that the relation between last-status-change and certificate-validity-start days is not straightforward.
Some of the CAs have even weekly distributions for both processes, which may suggest higher levels of automation (e.g., Let's Encrypt, Google, Actalis, cPanel, Gandi, Herndon). Among the large CAs, DigiCert stands out with their pronounced weekly patterns for both processes.  
Similarly, there are differences in the daily and hourly distributions of the expiry times selected for certificates (Figure~\ref{fig:expiry-heat}).  Here, some of the large CAs (e.g., Let's Encrypt, GoDaddy, Google, GlobalSign) spread expiry times both across the week and the hours of the days, whereas other large CAs (e.g., DigiCert, Comodo, cPanel, Sectigo) always set certificates to expire at the same time of day. These differences may not have major security implications, however, they demonstrate the lack of a standardized policy for managing the statuses of expired certificates.  

\section*{Appendix B: Responses from the CAs}

\surveyPAM{}{We contacted 8 organizations that operate the CAs for which we observed at least one status change 
from ``Revoked" to ``Good".
However, we did not find a contact email for one CA that no longer operates:
AT\&T Wi-Fi Services.  
We received responses from 5 organizations: Starfield (GoDaddy), Japan Registry, Entrust, ACCV, and Atos.
The CAs that responded confirmed that they had issued the certificates in question and provided varying explanations for their behavior. Two CAs argued that their use of ``Good" statuses was motivated by 
the standard~\cite{RFCOCSP}, which states that
``at a minimum, this positive response [i.e., a ``Good" response] indicates that no certificate with the requested certificate serial number currently within its validity interval is revoked." 
One of these two CAs also stated that they ``are going to consult with the community to clarify the requirements, and then, follow it."  We believe that CAs should avoid changing the status of revoked certificates to ``Good" at any time.}

\end{document}